\documentclass[prb,twocolumn,showpacs,amsmath,amssymb,superscriptaddress]{revtex4}
\usepackage{graphicx}
\usepackage{bm}

\begin{document}

\title{Reading-out the state inductively\\ and microwave spectroscopy
of an interferometer-type charge qubit}
\author{D. Born}
\affiliation{%
Friedrich Schiller University, Institute of Solid State Physics,
Helmholtzweg 5, D-07743 Jena, Germany}
\author {V.I. Shnyrkov}
\altaffiliation[On leave from ]{B. Verkin Institute for Low
Temperature Physics and Engineering, National Academy of Sciences
of the Ukraine, 310164 Kharkov, Ukraine.}
\affiliation{%
Friedrich Schiller University, Institute of Solid State Physics,
Helmholtzweg 5, D-07743 Jena, Germany}
\author{W.~Krech}
\email{owk@rz.uni-jena.de}
\affiliation{%
Friedrich Schiller University, Institute of Solid State Physics,
Helmholtzweg 5, D-07743 Jena, Germany}
\author{Th.\ Wagner}
\affiliation{%
Institute for Physical High Technology, P.O. Box 100239, D-07702
Jena, Germany}
\author{E. Il'ichev}
\affiliation{%
Institute for Physical High Technology, P.O. Box 100239, D-07702
Jena, Germany}
\author{M. Grajcar}
\altaffiliation[On leave from ]{Department of Solid State Physics,
Comenius University, SK-84248 Bratislava, Slovakia.}
\affiliation{%
Institute for Physical High Technology, P.O. Box 100239, D-07702
Jena, Germany}
\author {U. H\"ubner}
\affiliation{%
Institute for Physical High Technology, P.O. Box 100239, D-07702
Jena, Germany}
\author{H.-G. Meyer}
\affiliation{%
Institute for Physical High Technology, P.O. Box 100239, D-07702
Jena, Germany}

\date{\today}

\begin{abstract}
We implemented experimentally an interferometer-type charge qubit
consisting of a single-Cooper-pair transistor closed by a
superconducting loop that is in flip-chip configuration
inductively coupled to a radio frequency tank circuit. The tank
permits to readout the qubit state, acting as inductance measuring
apparatus. By applying continuous microwave power to the quantum
device, we observed inductance alterations caused by
redistributions of the energy level populations. From the measured
data, we extracted the energy gap between ground and upper levels
as a function of the transistor quasicharge as well as the
Josephson phase across both junctions.
\end{abstract}


\maketitle

Various quantum properties of superconducting structures with
small Josephson tunnel junctions have been experimentally
demonstrated by several research groups (see, e.g.,
Ref.~\onlinecite{Leggett} and references herein). The results have
been widely discussed in literature because such devices might
serve as prototypes for quantum bits (qubits).

In this context, the single-Cooper-pair transistor has attracted
renewed attention. The device consists of two mesoscopic Josephson
junctions coupled by a small island. The Hamiltonian of this
system can be written as\cite{ave}
\begin{equation}\label{ham}
 H = E_C(n-n_{g})^{2} - \varepsilon_{J}(\delta)
 \cos\varphi.
\end{equation}
Here the first term on the r.h.s. is the charging energy of the
island whereas the second one describes its Josephson coupling to
the leads. The variable $n=2m$ implies the number $m$ of excess
Cooper pairs on the central electrode, the parameter
$n_g=C_gV_g/e$ is continuously controllable by the gate voltage
$V_g$ via the capacitance $C_g$. The one-electron Coulomb energy,
$E_C=e^2/2C_\Sigma$, is expressed through the island's sum
capacitance $C_\Sigma$. Furthermore, the coupling strength,
\begin{equation}\label{jos}
\varepsilon_{J}(\delta)=[E_{J1}^{2}+E_{J2}^{2}+2E_{J1}E_{J2}\cos\delta]^{1/2},
\end{equation}
is a function of the total phase across both junctions,
$\delta=\varphi_1+\varphi_2$ (being here a good quantum variable).
$E_{J1,J2}$ are the individual Josephson energies and
$\varphi_{1,2}$ the respective junction phases. Note that the
observable $m$ is conjugated to the island's phase,
$\varphi=(\varphi_2-\varphi_1)/2$. In order to realize charge
qubits, we designed the system parameters to fulfill the
domination, $E_{CP}>\varepsilon_J(\delta)$, of the Cooper pair
Coulomb energy, $E_{CP}\equiv 4E_C$, over the coupling strength
$\varepsilon_J$.

In earlier studies, microwave induced transitions between ground
and upper energy states in single-charge transistors were observed
by measuring the switching current\cite{joy,eil,luk} or by means
of the photon-assisted quasiparticle cur\-rent\cite{Nak}. Later
the quantum coherent oscillations in similar configurations have
been detected by making use of microwave-pulsed
readouts.\cite{NakamuraPashkinTsai,Esteve}

Recently, a new possible solution was theoretically proposed by
several authors.\cite{Frie,Zorin} The main idea is to enclose a
single-Cooper-pair transistor into a superconducting inductive
loop forming an interferometer. The advantage of this circuit is
its low dissipation and, therefore, the remarkably weak
decoherence.\cite{Leggett} For realizing the quantum measuring
process, the loop is inductively coupled to a radio frequency tank
circuit.

\begin{figure}[t]
\includegraphics[width=8.0cm]{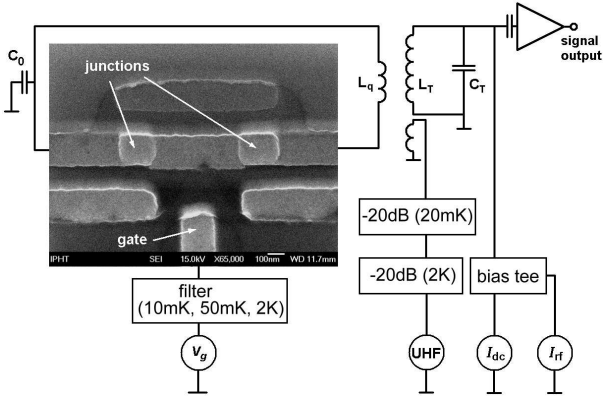}
\caption{Diagram of the measuring setup including the micrograph
of the Al based single-Cooper-pair transistor (closed by the
superconducting loop inductance $L_q$). On the r.h.s, the tank
circuit and some further electronic components are shown (cf.
text). The phase $\delta$ is guaranteed to be a good quantum
variable for sufficiently large ground capacitance, $C_0\gg
C_\Sigma$. The angular phase shift $\alpha$ (Eq.~\ref{alpha})
between tank voltage detected by the amplifier and current signal
$I_{rf}$ is determined by using a lock-in amplifier (not depicted
here).} \label{fig1}
\end{figure}

In this paper, we introduce experimentally the conception of an
interferometer-type charge qubit in conjunction with a readout
procedure managed by measuring the qubit's reverse Josephson
inductance,\cite{Zorin, Krech}
\begin{equation}\label{ind1}
{\cal L}_J^{-1}=(2e/\hbar)^2\cdot\partial^2 H/\partial\delta^2.
\end{equation}
We put it into practice along the lines of the conventional
impedance measuring technique\cite{Rifkin} by means of a
high-quality tank inductively coupled to the interferometer loop.
Based on this method, we present results of the spectroscopic
investigation of the quantum device that is manipulated by a
microwave field injected via a coaxial cable (UHF line, Fig.
\ref{fig1}). These results demonstrate the utility of the used
design relating to development and characterization (weak
continuous quantum measuring) of future building blocks (coupled
qubits) for quantum information circuits as well.

The principle of measurement is as follows: The expectation value
$\langle m|{\cal L}_J^{-1}(n_g,\delta)|m\rangle$ is determined not
only by the quasicharge $en_g$ (controlled by the gate voltage)
and the phase $\delta$ (controlled by an external magnetic flux
$\Phi_e$ threading the interferometer loop) but also by the band
index $m$.\cite{ind} Within the two-level approximation, the mean
values in the upper state ($m$=1) and in the ground state ($m$=0)
have the opposite sign,\cite{Krech} $\langle 1|{\cal
L}_J^{-1}(n_g,\delta)|1\rangle\simeq -\langle 0|{\cal
L}_J^{-1}(n_g,\delta)|0\rangle$. Consequently, a change of the
band index $m$ results in a relevant impedance change that causes
a shift of the tank resonance frequency.

Quantitatively, we consider only the case when the geometrical
inductance $L_q$ of the loop is sufficiently small, ensuring in
this way a unique relationship between the flux $\Phi_e$ and the
phase difference $\delta$. This requirement can be expressed as
follows within the two-band model, :\cite{Krech}
\begin{equation}\label{Lstar}
L_q<2\left(\frac{\Phi_0}{2\pi}\right)^2\frac{\Delta E(n_g=1,
\delta=\pi)}{E_{J1}E_{J2}},
\end{equation}
where $\Phi_0$ is the flux quantum and
\begin{equation}\label{erg}
\Delta E =\left[D^2(n_g)+\varepsilon_J^2(\delta )\right]^{1/2}
\end{equation}
is the local band spacing (with $D\equiv E_{CP}(1-n_g)$). We
emphasize the fact that the coupling strength $\varepsilon_J$ and,
therefore, the gap $\Delta E$ can be significantly reduced in the
vicinity of the operating point $n_g$=1, $\delta$=$\pi$ even for
relatively large ratios $E_{J1,J2} /E_C$.

As discussed above, the observable ${\cal L}_J^{-1}$ takes
different expectation values for the energy bands $m$=0 and $m$=1.
Therefore, a redistribution of the level populations caused by
microwave excitations (in competition with relaxation processes)
results in a change of the quantum-statistical mean value
${\langle\cal L}^{-1}_J\rangle$. In order to prove this effect,
the loop was coupled through a mutual inductance $M=k\sqrt{L_q
L_T}$ (where $k\ll 1$ is the coupling coefficient) to the tank
circuit (with known inductance $L_T$ and quality factor $Q$). The
tank is driven by an rf current $I_{rf}$ at a frequency $\omega$
close to its resonance frequency $\omega_T$. This current
generates an rf flux $\Phi_{rf}$ threading the interferometer
ring. Provided that, first, this rf flux is small,
$\Phi_{rf}\ll\Phi_0$, and, second, the condition $L_q|\langle{\cal
L}_J^{-1}\rangle|\ll 1$ is fulfilled, variations of the
interferometer inductance can be described by making use of the
formula\cite{Ilichev01}
\begin{equation}\label{alpha}
\tan\alpha\simeq k^2Q~L_q\langle{\cal L}_J^{-1}\rangle,
\end{equation}
where $\alpha$ is the phase shift between drive current $I_{rf}$
and tank voltage $U_T$ to be measured.

The used measurement setup is shown in Fig.~\ref{fig1}. The Nb
square-shaped pancake tank coil ($L_T\simeq 170$~nH) was
fabricated lithographically on an oxidized Si substrate. An
external capacitance $C_T$ is used to complete the resonance
circuit. This high-quality tank ($Q\simeq700$) is biased by the rf
current $I_{rf}$ at the resonance frequency of
$\omega_T/2\pi\simeq28$~MHz. The rf voltage $U_T$ across the tank
circuit was measured by means of a sequence of cold and room
temperature amplifiers. The angular phase shift $\alpha$ was
determined using an rf lock-in voltmeter.

\begin{figure}[t]
\includegraphics[width=8.0cm]{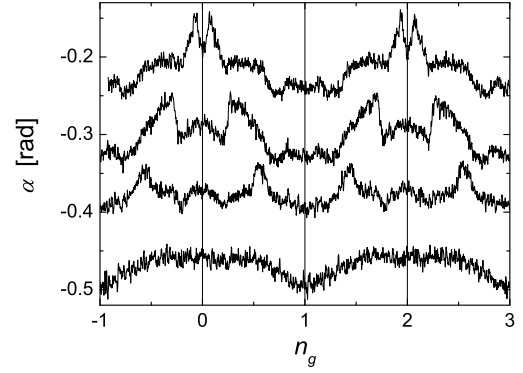}
\caption{Tank phase shift $\alpha$ vs. gate parameter $n_g$
without microwave power (lowest curve) and with microwave power at
different excitation frequencies. The data correspond to
$\omega_{UHF}/2\pi = 8.9, 7.5, 6.0$~GHz (from top to bottom). The
magnetic flux $\Phi_e$=$\Phi_0/2$ threading the interferometer
loop provides a total phase difference $\delta$=$\pi$ across the
single-Cooper-pair transistor. (For clarity, the upper curves are
shifted.)} \label{fig2}
\end{figure}

The qubit under investigation is placed in the middle of the coil
in flip-chip configuration. The transistor was fabricated out of
Al by the conventional shadow evaporation technique. The
transistor's gate line is filtered  by means of three copper
powder filters (with a total length of about 35~cm) at the
temperatures 2~K, 50~mK and 10~mK. The attenuation of this line is
about 120~dB at 5~GHz and 300~dB at 20~GHz. Microwave irradiation
for photon-assisted excitations of the quantum device is fed into
the sample via an UHF line consisting of a commercial coaxial
cable from room temperature to $\sim$2~K as well as via a
resistive coaxial cable (known as ThermoCoax) from $\sim$2~K to
10~mK. In order to reduce external interferences, two 20~dB
commercial attenuators were installed at the temperature levels
2~K and 10~mK. The external magnetic flux $\Phi_e$ applied to the
interferometer loop is produced by a dc current $I_{dc}$ through
the tank coil. In order to feed both currents $I_{dc}$ and
$I_{rf}$ into the tank, a simple bias tee is used (see
Fig.~\ref{fig1}).

\begin{figure}[t]
\includegraphics[width=8.0cm]{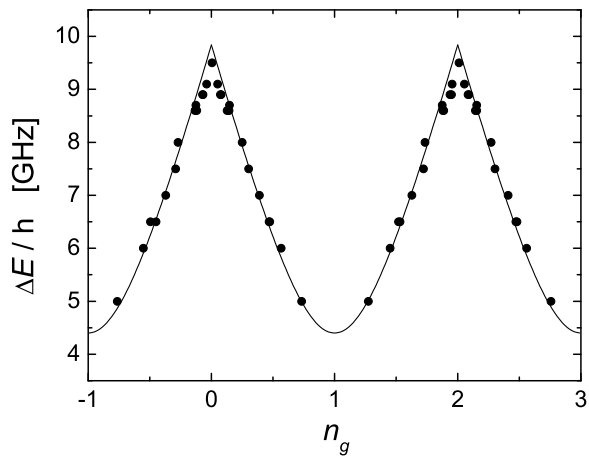}
\caption{Energy gap $\Delta E$ between the ground and upper bands
of the transistor determined from the experimental data for the
case $\delta$=$\pi$. (Some examples of these data are shown in
Fig.~\ref{fig2}.) Dots represent the experimental data, the solid
line corresponds to the fit within the two-band model (cf. text).}
\label{fig3}
\end{figure}

The angular phase shift $\alpha$ (\ref{alpha}) was measured as a
function of the gate voltage as well as of the dc current in the
tank coil that generates the external flux $\Phi_e$. The coupling
coefficient between tank and superconducting loop was obtained
experimentally\cite{Ilichev01} to be $k=0.026$. The mutual
inductance is $M=0.42$~nH, and the geometrical loop inductance was
calculated to be $L_q=1.5$~nH. Suitable adjusting the tank drive
current $I_{rf}$ permitted the qubit operation in the required
small-signal limit, $\Phi_{rf}\sim$~10$^{-3}\Phi_0$. The layout
size of the junctions is 140~nm$\times$180~nm. Deviations from
these nominal dimensions in the order of 10~$\%$ (caused by the
fabrication process) were estimated by means of the micrograph of
the real structure. In accordance with this fact, the
spectroscopically determined Josephson coupling energies showed a
certain difference, $|E_{J1}-E_{J2}|\approx 5$~GHz (see below).
The measurements were performed in a dilution refrigerator at a
nominal temperature of about 10~mK inside magnetic and
superconducting shields.

In a first set of experiments, we investigated charging effects in
the interferometer. Adjusting the external flux ($\Phi_e
=\Phi_0/2$), we fixed the phase ($\delta=\pi$) and measured the
dependence of the statistical average $\langle{\cal
L}_J^{-1}\rangle$ (or $\alpha$, cf. Eq.~\ref{alpha}) on the
parameter $n_g$. This dependence turned out to be rather weak due
to the large ratio between Josephson and charging energies.
Nevertheless, the coupling strength could be reduced down to
$\varepsilon_J(\delta =\pi )\approx 5$~GHz. Therefore, a variation
of $\langle{\cal L }_J^{-1}\rangle$ with $n_g$ was actually
measurable for operating values of $\delta$ in the vicinity of
$\pi$. Indeed, the 2e-periodic oscillations of $\langle{\cal
L}_J^{-1}\rangle$ or $\alpha$ with respect to the gate charge
$en_g=C_gV_g$, explicitly demonstrating its quasicharge character,
are clearly seen in Fig.~\ref{fig2} (bottom curve).\cite{Gotz}

\begin{figure}[t]
\includegraphics[width=8.0cm,height=7.0cm]{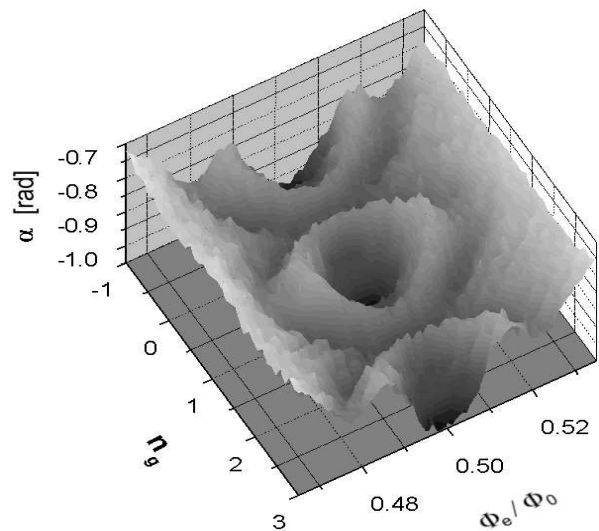}
\caption{Dependence of the phase shift $\alpha$ on the two
parameters $n_{g}$ and $\Phi_{e}$. The qubit is illuminated by a
microwave with the frequency of 8.0~GHz. The periodic circular
structure demonstrates the variation of the total
interferometer-tank impedance due to transitions from the lower to
the upper energy band. Especially, the ``crater ridges''
correspond to all combinations of the parameters $n_{g}$ and
$\Phi_{e}$ that give the same energy gap between the respective
states equal to 8.0~GHz.} \label{fig4}
\end{figure}

Small-amplitude continuous microwave irradiation of the sample
varied significantly the obtained dependence: Now distinguished
peaks appeared in the $\alpha(n_g)$ characteristics (upper curves
in Fig.~\ref{fig2}). Their positions depend on the frequency
$\omega_{UHF}$. Increasing microwave amplitudes led to a slight
change of the peak width only. Thus we demonstrated that the peaks
are originated by resonant excitations of the system from ground
to upper states. (In general, the measured peaks disappeared at
higher temperatures corresponding to the used frequency. For
instance, for a peak induced by the frequency $\omega_{UHF}/2\pi=
8$~GHz, this temperature is about $T \simeq 400$~mK.) The
microwave-induced transitions (at fixed parameters $\omega_{UHF}$
and $\delta$) from the ground to the upper states should
correspond to specific values $n_g$. Following this consideration
and extracting the peak positions from the $\alpha$-$n_g$ curves
measured at different microwave frequencies, we depicted the
energy differences $\Delta E=h\omega_{UHF}/2\pi$ between ground
and upper levels as a function of $n_g$ (Fig.~\ref{fig3}). Then we
fitted the spectroscopic data (analogously as reported in Ref. 8)
by making use of the band gap (\ref{erg}) within the two-level
approximation, providing $\varepsilon_J(\delta=\pi)=4.4$~GHz and
$E_C=2.2$~GHz. (A fitting procedure by making use of a numerical
treatment of the complete Hamiltonian (\ref{ham}) delivers
practically the same values.)

\begin{figure}[t]
\includegraphics[width=8.0cm,height=7.5cm]{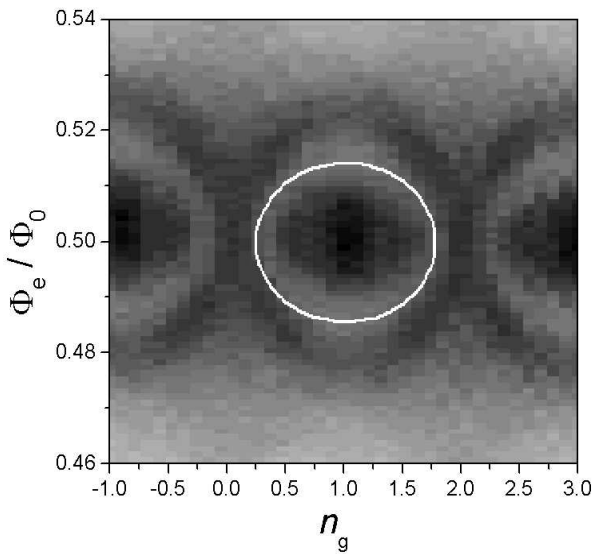}
\caption{Top view of the $\alpha (n_g,\Phi_e)$ mountain range of
Fig.~\ref{fig4}. Here the white ring approximates the ridge of the
central crater leading to the fit $E_C=2.2$~GHz,
$E_{J1}=70.0$~GHz, and $E_{J2}=74.4$~GHz within the two-level
model.}\label{fig5}
\end{figure}

In further experiments, we measured the phase shift $\alpha$ as a
function of the applied magnetic flux $\Phi_e$ in the vicinity of
$\Phi_0/2$ at varying gate voltage $V_g$. At the same time, the
microwave frequency was fixed. This way we obtained a spatial
representation of the microwave-induced transitions as a function
of both parameters $n_g$ and $\Phi_{e}$. The results for the
microwave frequency of 8~GHz are presented in Fig.~\ref{fig4}.
Finally fitting the ring-shaped ``crater rim'' by means of the
band spacing Eq.~\ref{erg}, we found spectroscopic data
$E_{J1}=70.0$~GHz and $E_{J2}=74.4$~GHz for the junctions'
Josephson energies (cf. Fig.~\ref{fig5}).

In conclusion, we have successfully implemented an alternative
quantum two-level system in a superconducting circuit with
readout. In this connection, we demonstrated the suitability of
the Josephson quantum inductance term for developing a readout
procedure. We performed a microwave spectroscopy of an
interferometer-type charge qubit, making use of an rf tank
readout. We reconstructed the energy gap between lower and upper
energy levels as a function of the gate charge at fixed phase
across the double junction. We demonstrated experimentally that
the energy gap can be varied by changing both gate charge and
external flux threading the interferometer loop. In view of the
relatively large ratio $E_J /E_C$, effects accompanied with the
change of the gate charge were observed in the vicinity of
$\delta$=$\pi$ only. Our results indicate that the investigated
qubit device is a potential candidate for realizing solid-state
quantum processing in conjunction with weak characterization.

We thank A.~Smirnov, A.~Zagoskin, and A.~Zaikin for fruitful
discussions. We also would like to thank A.N.~Omelyanchouk for
comments and H. M\"{u}hlig for technical assistance. This work was
partially supported by the Deutsche Forschungsgemeinschaft under
contract No. KR 1172/9-2 and D-wave Sys. Inc.


\begin{thebibliography}{99}
\bibitem{ind} The band index can be related simply to the number $m$
of excess pairs on the island in the pure charge states
($\varepsilon_J\rightarrow 0; 0<n_g<1$). Within the two-band
model, this procedure leads to $|m=0\rangle$ (lower state) and
$|m=1\rangle$ (upper state), respectively (cf. Ref. 7).
\end{thebibliography}
\end{document}